\documentclass{jnmp-arxiv}

\begin{document}


\newcommand{\eref}[1]{(\ref{#1})}

\newcommand{\mymatrix}[1]{\mathsf{#1}}
\newcommand{\mypmatrix}[1]{\begin{pmatrix}#1\end{pmatrix}}
\newcommand{\mySpd}[1]{\mathbb{T}_{#1}}
\newcommand{\mySpi}[1]{\mathbb{S}_{#1}}
\newcommand{\mySnd}[1]{\overline{\mathbb{T}}_{#1}}
\newcommand{\mySni}[1]{\overline{\mathbb{S}}_{#1}}
\newcommand{\myTheta}{c}
\newcommand{\mytheta}{\bar{c}}
%

\markboth{V.E. Vekslerchik}
{Functional representation of the negative DNLS hierarchy}

%
\catchline{}{}{2013}{}{}
%

\copyrightauthor{V.E. Vekslerchik}

\title{Functional representation of the negative DNLS hierarchy}

\author{V.E. Vekslerchik}

\address{
Usikov Institute for Radiophysics and Electronics \\
12, Proskura st., Kharkov, 61085, Ukraine \\
\email{vekslerchik@yahoo.com}
}

\maketitle

\vphantom{\vbox{%
\begin{history}
\received{(Day Month Year)}
\revised{(Day Month Year)}
\accepted{(Day Month Year)}
\end{history}
}}

\begin{abstract}
This paper is devoted to the negative flows of the derivative nonlinear 
Schr\"odinger hierarchy (DNLSH). The main result of this work is the 
functional representation of the extended DNLSH, composed of both positive 
(classical) and negative flows. We derive a finite set of functional equations, 
constructed by means of the Miwa's shifts, which contains all equations of 
the hierarchy. Using the obtained functional representation we convert the 
nonlocal equations of the negative subhierarchy into local ones of higher order, 
derive the generating function of the conservation laws and the N-soliton 
solutions for the extended DNLSH under non-vanishing boundary conditions.
\end{abstract}


\ccode{2010 Mathematics Subject Classification: 
  37J35, 
  35Q51, 
  35Q55, 
  37K10  
}

\section{Introduction. \label{sec-intro}}

This paper, which can be viewed as a continuation of \cite{V12,V11}, is devoted 
to  the negative flows of the derivative nonlinear Schr\"odinger (DNLS) 
hierarchy (DNLSH) \cite{KN78,CLL79,GI83}. 
Of three known forms of the DNLS equation 
(the Kaup-Newell \cite{KN78}, Chen-Lee-Liu \cite{CLL79} 
and Gerdjikov-Ivanov \cite{GI83} equations) 
we use as the starting point the Chen-Lee-Liu equation, 
\begin{equation}
  i U_{t} + U_{xx} \pm 2i |U|^{2}U_{x} = 0,
\end{equation}
which we write as a system
\begin{subequations}
\label{syst-dnlse}
\begin{eqnarray}
  i U_{t} + U_{xx} + 2i UVU_{x} & = & 0, 
\\
  -i V_{t} + V_{xx} + 2i UVV_{x} & = & 0. 
\end{eqnarray}
\end{subequations}
Consequently, we use the term DNLSH for the hierarchy whose simplest equations 
are \eref{syst-dnlse} 
(see the recent paper \cite{FGZ13} on the negative flows of the Kaup-Newell 
hierarchy).

In this work we address several problems. 
First, we want to describe the extended DNLSH composed of both positive and 
negative flows and our main result is the functional representation of the 
positive, negative and mixed DNLS subhierarchies 
(sections \ref{sec-pos}-\ref{sec-mix}). We derive a finite set of functional 
equations, constructed by means of the Miwa's shifts, which contains all 
equations of the extended hierarchy, that can be recovered by the power 
series expansion. 

Considering various limiting cases of the obtained equations we demonstrate 
in section \ref{sec-exm} that 
the extended DNLSH contains except the massive Thirring model 
(which is already known fact) such systems as Mikhailov-Fokas-Lenells equation, 
(1+2)-dimensional versions of the Chen-Lee-Liu and Kaup-Newell equations as well 
as a 2-dimensional elliptic system with Toda-like nonlinearity which differs, 
apparently, from the DNLS-like systems.

The second goal of this paper is to derive the generating function 
for the conservation laws, which is presented in section \ref{sec-cl}. 
Contrary to the approach based on the inverse scattering transform, 
where such functions are expressed in terms of the scattering data and 
Jost functions, we formulate all results explicitly, in terms of solutions of 
the DNLSH.

Finally, we obtain in sections \ref{sec-bilin} and \ref{sec-dark} the 
$N$-soliton solutions for the extended DNLSH under non-vanishing boundary 
conditions. 
Here we use another advantage of the functional representation and demonstrate 
how this problem can be solved by elementary algebraic calculations. 

\section{Holonomy representation of the extended DNLSH. \label{sec-hr}}

Instead of the so-called zero-curvature representation (ZCR) 
that is based on presenting the equations as the compatibility conditions 
for the linear \emph{differential} systems
\begin{equation}
  \frac{\partial}{\partial t_{j}} \Psi = \mymatrix{M}_{j} \Psi, 
  \qquad
  \frac{\partial}{\partial \bar{t}_{k}} \Psi = \widetilde{\mymatrix{M}}_{k} \Psi, 
  \qquad
  j,k = 1, 2, ...
\end{equation}
we deal with the linear systems constructed by the Miwa's shift 
operators that are applied to functions of a doubly infinite number of arguments,
\begin{equation}
  U( \mathrm{t}, \bar{\mathrm{t}} ) 
  = 
  U\left( t_{1}, t_{2}, ... , \bar{t}_{1}, \bar{t}_{2}, ... \right) 
  = 
  U\left( t_{j}, \bar{t}_{k} \right)_{j,k = 1, 2, ...}, 
\end{equation}
and are defined by
\begin{equation}
\begin{array}{lcl}
  \mySpd{\xi} U\left( \mathrm{t}, \bar{\mathrm{t}} \right) 
  & = & 
  U\left( \mathrm{t} + i[\xi], \bar{\mathrm{t}} \right),
\\
  \mySnd{\eta} U\left( \mathrm{t}, \bar{\mathrm{t}} \right) 
  & = & 
  U\left( \mathrm{t}, \bar{\mathrm{t}} + i [\eta] \right) 
\end{array}
\end{equation}
or 
\begin{equation}
\begin{array}{lcl}
  \mySpd{\xi} 
  U\left( t_{j}, \bar{t}_{k} \right)_{j,k = 1, 2, ...} 
  & = & 
  U\left( t_{j} + i \xi^{j} / j, \bar{t}_{k} \right)_{j,k = 1, 2, ...}, 
\\
  \mySnd{\eta} 
  U\left( t_{j}, \bar{t}_{k} \right)_{j,k = 1, 2, ...} 
  & = & 
  U\left( t_{j}, \bar{t}_{k} + i \eta^{k} / k \right)_{j,k = 1, 2, ...}
\end{array}
\end{equation}
Thus, our goal is to study the compatibility conditions of the systems of the 
following type:
\begin{subequations}
\begin{eqnarray}
  \mySpd{\xi} \psi 
  & = & 
  \mymatrix{L}(\xi) \psi
\\
  \mySnd{\eta} \psi 
  & = & 
  \bar{\mymatrix{L}}(\eta) \psi,
\end{eqnarray}
\end{subequations}
(where $\psi$ is a $2$-vector, $\mymatrix{L}$ and $\bar{\mymatrix{L}}$ are 
$2 \times 2$-matrices) that are given by 
\begin{subequations}
\label{commutativity}
\begin{eqnarray}
  \left[\mySpd{\xi_{1}}\mymatrix{L}\left(\xi_{2}\right)\right] 
  \mymatrix{L}\left(\xi_{1}\right) 
  & = & 
  \left[\mySpd{\xi_{2}} \mymatrix{L}\left(\xi_{1}\right) \, \right] 
  \mymatrix{L}\left(\xi_{2}\right), 
\label{commutativity-PP}
\\[2mm]
  \left[ \, \mySpd{\xi} \bar{\mymatrix{L}}\left(\eta\right) \, \right] 
  \mymatrix{L}\left(\xi\right) 
  & = & 
  \left[ \, \mySnd{\eta}{} \mymatrix{L}\left(\xi\right) \, \right] 
  \bar{\mymatrix{L}}\left(\eta\right), 
\label{commutativity-PN}
\\[2mm]
  \left[ \, 
    \mySnd{\eta_{1}} \bar{\mymatrix{L}}\left(\eta_{2}\right) 
  \, \right] 
  \bar{\mymatrix{L}}\left(\eta_{1}\right) 
  & = & 
  \left[ \, 
    \mySnd{\eta_{2}} \bar{\mymatrix{L}}\left(\eta_{1}\right) 
  \, \right] 
  \bar{\mymatrix{L}}\left(\eta_{2}\right). 
\label{commutativity-NN}
\end{eqnarray}
\end{subequations}
Using the standard strategy of the ZCR, that consists in introducing an 
auxiliary parameter, $\zeta$, and looking for the matrices $\mymatrix{L}$ 
and $\bar{\mymatrix{L}}$ with simplest dependence on $\zeta$, one can obtain 
the matrices that lead to the DNLSH. It turns out that $\mymatrix{L}$ and 
$\bar{\mymatrix{L}}$ should be \textit{linear} functions of $\zeta$ and 
$\zeta^{-1}$ (see \cite{V12} for explanation). Omitting the details of the 
calculations we present here the `minimal' solution for \eref{commutativity}: 
\begin{equation}
\label{zcr-mat-pos}
  \mymatrix{L}(\xi) = 
  \mymatrix{L}(\xi; \zeta) = 
  \frac{ 1 }{ G(\xi) } \; 
  \mypmatrix{ \zeta\xi + G(\xi) & \quad\xi V \cr - \zeta\xi\mySpd{\xi} U & 1 } 
\end{equation}
and 
\begin{equation}
\label{zcr-mat-neg}
  \bar{\mymatrix{L}}(\eta) = 
  \bar{\mymatrix{L}}(\eta; \zeta) = 
  \mypmatrix{ \zeta^{-1}\eta + g(\eta) & \quad\zeta^{-1}\eta v \cr -\eta\mySnd{\eta} u & 1 } 
\end{equation}
where
\begin{eqnarray}
  G(\xi) & = & 1 + \xi (\mySpd{\xi}U) V, 
\label{gp-def} 
\\
 g(\eta) & = & 1 + \eta (\mySnd{\eta} u) v. 
\label{gn-def} 
\end{eqnarray}
The functions $U$, $V$, $u$ and $v$ are subjected to the constraints that 
are discussed below. 

In the following sections we study the systems of equations 
(the subhierarchies of the extended DNLSH) that appear as the result 
of \eref{commutativity} combined with \eref{zcr-mat-pos} and \eref{zcr-mat-neg}.

\section{Positive DNLS subhierarchy. \label{sec-pos}}

The equations that follow from \eref{commutativity-PP} are 
\begin{subequations}
\label{pp-miwa} 
\begin{eqnarray}
  0 & = &
  \left( \xi_{1} G_{2} - \xi_{2} G_{1} \right)  (\mySpd{12}U) 
  - \xi_{1} (\mySpd{1}U) 
  + \xi_{2} (\mySpd{2}U) 
\label{pp-miwa-u} 
\\
  0 & = &
  \left[ \xi_{1} (\mySpd{1}G_{2}) - \xi_{2} (\mySpd{2}G_{1}) \right]  V 
  - \xi_{1} (\mySpd{2}V) 
  + \xi_{2} (\mySpd{1}V) 
\label{pp-miwa-v} 
\end{eqnarray}
\end{subequations}
where we use the simplified notation
\begin{equation}
  \mySpd{j} = \mySpd{\xi_{j}}, 
  \qquad
  G_{j} = G(\xi_{j}), 
  \qquad
  j=1,2.
\end{equation}
By simple algebra one can verify that equations \eref{pp-miwa} ensure 
vanishing of all components of the matrix equation \eref{commutativity-PP}. 
This system, that can be viewed as the functional representation of the 
positive DNLSH, has been derived in \cite{V02,DM06}.

Equations \eref{pp-miwa} can be viewed as `basic' representation from which one 
can obtain a few different systems that are more appropriate for tackling 
different problems.

For example, sending $\xi_{2}$ to zero one arrives at the system 
\begin{subequations}
\label{pp-part1} 
\begin{eqnarray}
  i \xi \;\partial_{1}U  
  & = & 
  \left[ 1+ \xi U (\mySpi{\xi}V) \right]  \left( U - \mySpi{\xi}U \right)  
\label{pp-part1-u} 
\\
  i \xi \;\partial_{1}V  
  & = &
  \left[ 1+ \xi (\mySpd{\xi}U) V \right]  \left( \mySpd{\xi}V - V \right)  
\label{pp-part1-v} 
\end{eqnarray}
\end{subequations}
where 
\begin{equation}
  \mySpi{} = \mySpd{}^{-1} 
\end{equation}
and $\partial_{1} = \partial/\partial t_{1}$. 
From this representation it is easy to derive the `individual' equations of 
the hierarchy. The simplest of them, which is the result of 
collecting the $\xi^{2}$-terms, is \eref{syst-dnlse}.

Another form of \eref{pp-miwa} can be obtained by introducing the operator 
$\partial(\xi)$ by 
\begin{equation}
  \partial(\xi) = \sum_{j=1}^{\infty} \xi^{j} \partial_{j}. 
\label{part-pos}
\end{equation}
Noting that
\begin{equation}
  \lim\limits_{ \xi_{1},\xi_{2} \to \xi } \; 
  \frac{ 1 }{ \xi_{1} - \xi_{2} } \; 
  \left( 
    \mySpd{\xi_{1}}\mySpd{\xi_{2}}^{-1} - 1 
  \right) f 
  = 
  i \xi^{-1} \partial(\xi) \, f 
\label{def-part-pos}
\end{equation}
one can rewrite \eref{pp-miwa} as 
\begin{subequations}
\label{pp-part-xi}
\begin{eqnarray}
  i \;\partial(\xi)U  
  & = & 
  H(\xi) \left( \mySpd{\xi}U - U \right), 
\\
  - i \;\partial(\xi)V  
  & = & 
  H(\xi) \left( \mySpi{\xi}V - V \right)  
\end{eqnarray}
\end{subequations}
with 
\begin{equation}
  H(\xi) 
  = 
  \left [ 1  + \xi (\mySpd{\xi}U) (\mySpi{\xi}V) \right]^{-1}.
\label{wp-def} 
\end{equation}
We use this representation to derive the the generating function 
for the conservation laws (see section \ref{sec-cl}).

Bearing in mind the bilinearization that we need to obtain explicit solutions 
for the DNLSH it is convenient to introduce the `potential' $\Psi$. 
Noting that 
\begin{equation}
  \left[\mySpd{\xi_{2}}G(\xi_{1})\right] G(\xi_{2}) 
  - G(\xi_{1}) \left[\mySpd{\xi_{1}}G(\xi_{2})\right] 
  = 
  0
\label{pp-gg} 
\end{equation}
(this identity can be proved by straightforward application of \eref{pp-miwa}) 
one can present $G(\xi)$  as 
\begin{equation}
  G(\xi) 
  = 
  \Phi (\mySpd{\xi}\Psi) 
\label{gp-psi} 
\end{equation}
where 
\begin{equation}
  \Phi = \Psi^{-1} 
\end{equation}
and rewrite \eref{pp-miwa} as follows:
\begin{subequations}
\begin{eqnarray}
  \xi_{1}^{-1} 
  \left( 1 - \mySpi{1} \right) \left(\mySpi{2}\Psi \right) U 
  & = & 
  \xi_{2}^{-1} 
  \left( 1 - \mySpi{2} \right) \left(\mySpi{1}\Psi \right) U, 
\\
  \xi_{1}^{-1} 
  \left( 1 - \mySpd{1} \right) \left(\mySpd{2}\Phi \right) V 
  & = & 
  \xi_{2}^{-1} 
  \left( 1 - \mySpd{2} \right) \left(\mySpd{1}\Phi \right) V. 
\end{eqnarray}
\end{subequations}

The positive DNLS subhierarchy, to repeat, is the classical DNLS hierarchy, 
that has been introduced in the 70's and which is one of the most well-studied 
integrable systems. That is why we do not discuss equations \eref{pp-miwa} 
and their consequences here in detail. 

\section{Negative DNLS subhierarchy. \label{sec-neg}}

Equations that follow from the commutativity condition \eref{commutativity-NN} 
can be written as 
\begin{subequations}
\label{nn-miwa}
\begin{eqnarray}
 0  
 & = & 
 \left( \eta_{1} g_{{2}} 
- \eta_{2} g_{{1}} \right) (\overline{\mathbb{T}}_{12} u) 
- \eta_{1} (\overline{\mathbb{T}}_{1} u) 
+ \eta_{2} (\overline{\mathbb{T}}_{2} u), 
\\
 0 
 & = & 
 \left( \eta_{1} (\overline{\mathbb{T}}_{1} g_{{2}}) 
- \eta_{2} (\overline{\mathbb{T}}_{2} g_{{1}}) \right) v 
- \eta_{1} (\overline{\mathbb{T}}_{2} v) 
+ \eta_{2} (\overline{\mathbb{T}}_{1} v). 
\end{eqnarray}
\end{subequations}
It is easy to see that these equations become nothing but equations \eref{pp-miwa}
after the substitution
\begin{equation}
  u,v,\mySnd{} \to U,V,\mySpd{} 
\end{equation}
or, in other words, the `purely negative' DNLS subhierarchy is identical to 
the `classical' one. Thus, we do not repeat the consideration of the previous 
section and present here only the identity
\begin{equation}
 (\overline{\mathbb{T}}_{2} g_{{1}}) g_{{2}} 
- g_{{1}} (\overline{\mathbb{T}}_{1} g_{{2}}) 
  = 0 
\label{nn-gg}
\end{equation}
which is used below.

\section{Mixed DNLS subhierarchy. \label{sec-mix}}

In this section we study the main object of this paper, the mixed (or extended) 
DNLSH. 
The results of the previous section demonstrate that this hierarchy can be 
thinked of as a result of \emph{integrable coupling} of two copies of the 
`classical' DNLSH.
Equations that are discussed below are closely related to the equations that 
usually appear in the works devoted to the negative flows of any hierarchy. 

Substituting matrices \eref{zcr-mat-pos} and \eref{zcr-mat-neg} into 
\eref{commutativity-PN} one arrives at 
\begin{subequations}
\label{pn-miwa} 
\begin{eqnarray}
  g(\eta) (\mySnd{\eta}U) 
  - U 
  & = & 
  \eta (\mySnd{\eta}u) 
\\ 
  g(\eta) V 
  - \mySnd{\eta}V 
  & = & 
  \eta v 
\\[2mm]
  G(\xi) (\mySpd{\xi}u) 
  - u 
  & = & 
  \xi (\mySpd{\xi}U) 
\\ 
  G(\xi) v 
  - \mySpd{\xi}v 
  & = & 
  \xi V 
\end{eqnarray}
\end{subequations}
Using \eref{gp-psi} and similar representation of $g(\eta)$,  
\begin{equation}
  g(\eta) 
  = 
  \Psi (\mySnd{\eta}\Phi) 
\label{gn-psi} 
\end{equation}
which follows from \eref{nn-gg} and the identity 
\begin{equation}
  \left[\mySpd{\xi}g(\eta)\right] \left[\mySnd{\eta}G(\xi)\right] 
  = 
  g(\eta) G(\xi) 
\end{equation}
stemming from \eref{pp-miwa}, \eref{nn-miwa} and \eref{pn-miwa},  
one can eliminate from this system the `negative variables' $u$ and $v$ coming 
to equations that can be written  as
\begin{subequations}
\label{pn-upvp}
\begin{eqnarray}
  \left[ \mySpi{\xi} - (\mySpi{\xi}\Phi) \Psi \right] 
  \left[  \mySni{\eta}   - \Phi (\mySni{\eta}\Psi) \right] U  
  & = & 
  \xi \eta U, 
\\ 
  \left[ \mySpd{\xi} - \Phi (\mySpd{\xi}\Psi) \right] 
  \left[ \mySnd{\eta} - (\mySnd{\eta}\Phi) \Psi \right] V  
  & = & 
  \xi \eta V. 
\end{eqnarray}
\end{subequations}
Namely this system may be viewed as the mixed, or extended, DNLSH 
because it is written for the functions $U$ and $V$ (for which the equations 
of the positive DNLSH are written) and describes both positive and negative 
flows. It is still non-local because of the presence of $\Psi$, but can be 
made such, if considered together with, e.g. \eref{gp-psi}, 
\begin{equation}
  \mySpd{\xi}\Psi 
  = 
  \left[ 1 + \xi (\mySpd{\xi}U) V \right]\Psi , 
\end{equation}
as a system for the triple $U$, $V$ and $\Psi$.
However, it is more convenient to work with the `first-order' system 
\eref{pn-miwa}, which we rewrite now in a few alternative forms.

Passing from the Miwa shifts $\mySpd{\xi}$ and $\mySnd{\eta}$ to the 
$\partial(\xi)$- and $\bar\partial(\eta)$-operators, where $\partial(\xi)$ is defined by  
\eref{part-pos} and 
\begin{equation}
  \bar\partial(\eta) = 
  \sum_{k=1}^{\infty} \eta^{k} \bar\partial_{k}, 
  \qquad
  \bar\partial_{k} = \partial/\partial \bar{t}_{k},
\label{part-neg}
\end{equation}
one can obtain 
\begin{subequations}
\label{pn-part}
\begin{eqnarray}
  i \;\bar\partial(\eta)U  
  & = & 
  \eta h(\eta) \left[ 1- U (\mySni{\eta}v) \right]  (\mySnd{\eta}u) 
\\ 
  - i \;\bar\partial(\eta)V  
  & = & 
  \eta h(\eta) \left[ 1- (\mySnd{\eta}u) V \right]  (\mySni{\eta}v) 
\\[2mm]
  i \;\partial(\xi)u  
  & = & 
  \xi H(\xi) \left[ 1- u (\mySpi{\xi}V) \right]  (\mySpd{\xi}U) 
\label{eq:1811} 
\\
  - i \;\partial(\xi)v  
  & = & 
  \xi H(\xi) \left[ 1- (\mySpd{\xi}U) v \right]  (\mySpi{\xi}V) 
\label{eq:1812} 
\end{eqnarray}
\end{subequations}
where $H(\xi)$ is given by \eref{wp-def} and 
\begin{equation}
  h(\eta)
  = 
  \left[ 1 + \eta (\mySnd{\eta}u) (\mySni{\eta}v) \right]^{-1}
\end{equation}
Calculating the derivatives of $G(\xi)$ and $g(\eta)$ and comparing the result 
with \eref{gp-psi} and \eref{gn-psi} one arrives at
\begin{subequations}
\begin{eqnarray}
  H(\xi) 
  & = & 
  1 - i \;\partial(\xi) \ln\Psi  
\\
  h(\eta) 
  & = & 
  1 + i \;\bar\partial(\eta)\ln\Psi  
\end{eqnarray}
\end{subequations}
which leads to, probably, the shortest form of the extended DNLSH:  
\begin{subequations}
\label{pn-part-psi}
\begin{eqnarray}
  i \;\bar\partial(\eta) \, \Phi U    
  & = & 
  \eta h(\eta) \Phi (\mySnd{\eta}u) 
\\
  - i \;\bar\partial(\eta) \, \Psi V 
  & = & 
  \eta h(\eta) \Psi (\mySni{\eta}v) 
\\[2mm]
  i \;\partial(\xi) \, \Psi u 
  & = & 
  \xi H(\xi) \Psi (\mySpd{\xi}U) 
\\
  - i \;\partial(\xi) \, \Phi v 
  & = & 
  \xi H(\xi) \Phi (\mySpi{\xi}V) 
\end{eqnarray}
\end{subequations}

\subsection{Examples. \label{sec-exm}}

Here we present a few simplest (and hence most representative) equations of the 
mixed DNLSH.

\subsubsection{Example 1: Mikhailov-Fokas-Lenells equation.} 

One can obtain from \eref{pn-part-psi} in the $\xi,\eta \to 0$ limit that 
functions 
\begin{equation}
  Q = \Phi U, 
  \qquad 
  R = \Psi V 
 \label{ex-QR}
\end{equation}
considered as functions of 
\begin{equation}
  x=t_{1}, 
  \qquad
  y=\bar{t}_{1} 
\end{equation}
satisfy 
\begin{subequations}
\begin{eqnarray}
  0
  & = & 
  Q_{xy}  
  - 2 i Q R \;Q_{y}  
  + Q, 
\\
  0
  & = & 
  R_{xy}  
  + 2 i Q R \;R_{y}  
  + R 
\end{eqnarray}
\end{subequations}
where the subscripts stand for the derivatives with respect to the corresponding 
variables. 
It is easy to see that these equations are the 
relativistically invariant two-dimensional field model studied by Mikhailov 
(see \cite{GIK80,GI83,FGZ13}) which has reappeared in the recent literature as 
the Fokas-Lenells system \cite{F95,L09,LF09}.

\subsubsection{Example 2: (1+2)-dimensional Chen-Lee-Liu equation.}%

Considering equations \eref{pn-part} or \eref{pn-part-psi} in the $\xi \to 0$ 
limit one can obtain for the functions $Q$ and $R$ given by \eref{ex-QR} 
and function $P(\eta)$, 
\begin{equation}
  P(\eta) 
  = 
  h(\eta) 
  - 1
\label{ex-p-eta} 
\end{equation}
the following system:
\begin{subequations}
\begin{eqnarray}
  0
  & = & 
  \bar\partial_{1}\bar\partial(\eta)Q  
  + i \eta^{-1} \bar\partial(\eta)Q  
  - i \left[ 1+ 2 P(\eta) \right] \bar\partial_{1}Q, 
\\
  0
  & = & 
  \bar\partial_{1}\bar\partial(\eta)R  
  - i \eta^{-1} \bar\partial(\eta)R  
  + i \left[ 1+ 2 P(\eta) \right] \bar\partial_{1}R,  
\\
  0 
  & = & 
  \partial_{1}P(\eta)  
  - \bar\partial(\eta) Q R 
\end{eqnarray}
\end{subequations}
which after expansion in power series in $\eta$ leads to 
\begin{subequations}
\begin{eqnarray}
  0
  & = & 
  \bar\partial_{1}\bar\partial_{k} Q  
  + i \bar\partial_{k+1}Q  
  - 2i P^{(k)} \;\bar\partial_{1}Q, 
\\
  0
  & = & 
  \bar\partial_{1}\bar\partial_{k} R  
  - i \bar\partial_{k+1} R  
  + 2i P^{(k)} \;\bar\partial_{1}R, 
\\
  0 
  & = & 
  \partial_{1} P^{(k)}  
  - \bar\partial_{k}  Q R 
\end{eqnarray}
\end{subequations}
(here, $P^{(k)}$ are the coefficients of the Taylor series for $P(\eta)$).
The simplest equations of this hierarchy, rewritten in terms of variables 
$t$, $x$ and $y$, 
\begin{equation}
\label{ex-cll-times}
  t=\bar{t}_{2}, 
  \qquad
  x=\bar{t}_{1}, 
  \qquad
  y=t_{1},
\end{equation}
are
\begin{subequations}
\label{ex-cll-21}
\begin{eqnarray}
  0
  & = & 
  \phantom{+} i Q_{t}  
  + Q_{xx}  
  - 2i P Q_{x}, 
\\
  0
  & = & 
  - i R_{t}  
  + R_{xx}  
  + 2i P R_{x}, 
\\
  0 
  & = & 
  P_{y}  
  - (Q R)_{x} 
\end{eqnarray}
\end{subequations}
with $P = P^{(1)}$.
It is easy to see that the reduction $x=y$ converts equations \eref{ex-cll-21} 
into the Chen-Lee-Liu equation \eref{syst-dnlse}, thus one can consider them 
as a (1+2)-dimensional version of the latter.

\subsubsection{Example 3: (1+2)-dimensional Kaup-Newell equation.} 

Proceeding as in the previous example, but choosing this time 
\begin{equation}
  Q = \Psi U, 
  \qquad
  R = \Phi V 
\end{equation}
with $P(\eta)$ being defined by \eref{ex-p-eta}, 
one can derive from \eref{pn-part}
\begin{subequations}
\begin{eqnarray}
  0
  & = & 
  \partial_{1} \left[ \bar\partial(\eta) + 2 i P(\eta) \right] Q  
  + i \eta \;\bar\partial(\eta)Q  
  + \eta Q, 
\\
  0
  & = & 
  \partial_{1} \left[ \bar\partial(\eta) - 2 i P(\eta) \right] R   
  - i \eta \;\bar\partial(\eta)R  
  + \eta R, 
\\
  0 
  & = & 
  \partial_{1}P(\eta) - \bar\partial(\eta) Q R. 
\end{eqnarray}
\end{subequations}
The simplest of these equations (which come from collecting the $\eta^{2}$-terms) 
can be presented in terms of 
\begin{equation}
\label{ex-kn-times}
  P = P^{(2)}, 
  \qquad
  t = \bar{t}_{1}, 
  \qquad
  x = t_{1},
  \qquad
  y =\bar{t}_{2} 
\end{equation}
as the system 
\begin{subequations}
\label{ex-kn-21}
\begin{eqnarray}
  0
  & = & 
  \phantom{+} i Q_{t}  
  + Q_{xy}  
  + 2i (P Q)_{x}, 
\\
  0
  & = & 
  - i R_{t}  
  + R_{xy}  
  - 2i (PR)_{x}, 
\\
  0 
  & = & 
  P_{x}  
  - (Q R)_{y} 
\end{eqnarray}
\end{subequations}
that is nothing but a (1+2)-dimensional version of the Kaup-Newell equation 
\cite{KN78}.

\subsubsection{Example 4: Adler-Shabat $H_{5}$ system.}
%
The appearance of Chen-Lee-Liu- and Kaup-Newell-like equations in the above 
examples is quite natural.
However, from the extended DNLSH one can `extract' some equations that are, 
at least seemingly, not of the DNLS-like type. One of them we want to present 
here.

It can be shown that equations \eref{pn-part} imply that the functions 
\begin{equation}
  \lambda = \ln U, 
  \qquad 
  \mu = \ln u
\end{equation}
satisfy 
\begin{subequations}
\label{ex-as} 
\begin{eqnarray}
  0 
  & = & 
  \partial_{1}\bar\partial_{1} \lambda 
  - i e^{\mu-\lambda} \partial_{1} \lambda 
  + i e^{\lambda-\mu} \bar\partial_{1} \lambda, 
\\
  0 & = & 
  \partial_{1}\bar\partial_{1} \mu 
  + i e^{\mu-\lambda} \partial_{1} \mu 
  - i e^{\lambda-\mu} \bar\partial_{1} \mu 
\end{eqnarray}
\end{subequations}
which is (after the redefinition of the $t_{1}$ and $\bar{t}_{1}$ variables) 
the $H_{5}$ system from the classification of Adler and Shabat \cite{AS06}.

\subsubsection{Example 5: Massive Thirring model.} 
%
Writing the simplest equations of \eref{pn-part}
\begin{equation}
  \begin{array}{rcl}
  i \;\bar\partial_{1}U  
  & = &
  - u v U + u, 
  \\[1mm]
  - i \;\bar\partial_{1}V  
  & = &
  - u v V + v, 
  \end{array}
\qquad
  \begin{array}{rcl}
  i \;\partial_{1}u  
  & = &
  - U V u + U, 
  \\[1mm]
  - i \;\partial_{1}v  
  & = &
  - U V v 
  + V 
  \end{array}
\end{equation}
or, in the case of the reduction $V = \mp U^{*}$, $v = \mp u^{*}$ where ${}^{*}$ 
stands for the complex conjugation, 
\begin{equation}
  \begin{array}{rcl}
  i \;\bar\partial_{1}U  
  & = &
  \pm |u|^{2} U + u, 
  \\[1mm]
  i \;\partial_{1}u  
  & = &
  \pm |U|^{2} u + U 
  \end{array}
\end{equation}
we reproduce results of \cite{GIK80,GI83,NCQL83,FGZ13} on the relationship between the DNLSH and 
the massive Thirring model \cite{M76,KM77,KN77,M79}.

One can generalize this proceedings by rewriting equations \eref{pn-part} 
in terms of the functions 
\begin{equation}
  \begin{array}{lcl}
  Q(\xi) 
  & = &
  H(\xi) (\mySpd{\xi}U), 
  \\[1mm]
  R(\xi) 
  & = &
  H(\xi) (\mySpi{\xi}V) 
  \end{array}
\end{equation}
and 
\begin{equation}
  \begin{array}{lcl}
  q(\eta) 
  & = &
  h(\eta) (\mySnd{\eta}u), 
  \\[1mm]
  r(\eta) 
  & = &
  h(\eta) (\mySni{\eta}v). 
  \end{array}
\end{equation}
Replacing $H(\xi)$ and $h(\eta)$ with 
\begin{equation}
  K(\xi) = 2 H(\xi) - 1, 
\qquad
  k(\eta) = 2 h(\eta) - 1
\end{equation}
which are related to $Q$, $R$, $q$ and $r$ by 
\begin{eqnarray}
  K(\xi) 
  & = & 
  \sqrt{ 1 - 4 \xi Q(\xi) R(\xi) }, 
\\[2mm]
  k(\eta) 
  & = & 
  \sqrt{ 1 - 4 \eta q(\eta) r(\eta) }
\end{eqnarray}
one can transform \eref{pn-part} into 
\begin{subequations}
\label{Thirring-pn} 
\begin{eqnarray}
  i \left( 1- \xi \eta \right)  \;\bar\partial(\eta)Q(\xi)  
  & = & 
  a(\xi,\eta) Q(\xi) 
  + \eta K(\xi) q(\eta), 
\\
  - i \left( 1- \xi \eta \right)  \;\bar\partial(\eta)R(\xi)  
  & = & 
  a(\xi,\eta) R(\xi) 
  + \eta K(\xi) r(\eta), 
\\[2mm] 
  i \left( 1- \xi \eta \right)  \;\partial(\xi)q(\eta)  
  & = & 
  A(\xi,\eta) q(\eta) 
  + \xi k(\eta) Q(\xi), 
\\
  - i \left( 1- \xi \eta \right)  \;\partial(\xi)r(\eta)  
  & = & 
  A(\xi,\eta) r(\eta) 
  + \xi k(\eta) R(\xi) 
\end{eqnarray}
\end{subequations}
where
\begin{eqnarray}
  2 A(\xi,\eta) 
  & = & 
  \left( 1+ \xi \eta \right)  K(\xi) 
  - 1 + \xi \eta, 
\\
  2 a(\xi,\eta) 
  & = & 
  \left( 1+ \xi \eta \right)  k(\eta) 
  - 1 + \xi \eta. 
\label{eq:2038} 
\end{eqnarray}
Equations \eref{Thirring-pn}, whose vector form is given by 
\begin{subequations}
\label{Thirring-vec} 
\begin{eqnarray}
  i \left( 1- \xi \eta \right)  \sigma_{3}\bar\partial(\eta)\mathbf{Q}(\xi)  
  & = & 
    a(\xi,\eta) \mathbf{Q}(\xi) 
  + \eta K(\xi) \mathbf{q}(\eta), 
\\[2mm] 
  i \left( 1- \xi \eta \right)  \sigma_{3}\partial(\xi)\mathbf{q}(\eta)  
  & = & 
  A(\xi,\eta) \mathbf{q}(\eta) 
  + \xi k(\eta) \mathbf{Q}(\xi) 
\end{eqnarray}
\end{subequations}
with 
\begin{equation}
  \mathbf{Q}(\xi) = \mypmatrix{ Q(\xi) \cr R(\xi) }, 
  \qquad
  \mathbf{q}(\eta) = \mypmatrix{ q(\eta) \cr r(\eta) }, 
  \qquad
  \sigma_{3} = \left( \begin{array}{lc} 1 & 0 \\ 0\;  & -1 \end{array}\right)  
\end{equation}
can be viewed as the massive Thirring hierarchy.

\section{Constants of motion and conservation laws. \label{sec-cl}}

The `classical' (positive) DNLSH, as an integrable system, possesses an 
infinite number of constants of motion that can be presented as 
\begin{equation}
  \mathcal{I}_{\ell}\left(t_{2},t_{3},...\right)
  = 
  \int \mathcal{J}_{\ell}\left(t_{1}, t_{2},t_{3},...\right) dt_{1},
  \qquad 
  \ell=0,1,...
\end{equation}
Clearly, one can consider $\mathcal{I}_{\ell}$ as functions of a twice infinite 
set of variables
\begin{equation}
  \mathcal{I}_{\ell} = 
  \mathcal{I}_{\ell}\left( t_{2}, t_{3}, ... , \bar{t}_{1}, \bar{t}_{2}, ... \right) 
\end{equation}
being the constants with respect to both positive and negative `times':
\begin{subequations}
\begin{eqnarray}
	\partial \mathcal{I}_{\ell} / \partial t_{j} = 0
	&\qquad&
	j=2,3,...
\\	
	\partial \mathcal{I}_{\ell} / \partial \bar{t}_{k} = 0
	&&
	k=1,2,...
\end{eqnarray}
\end{subequations}
It turns out that the generating function for $\mathcal{J}_{\ell}$,
\begin{equation}
  \mathcal{J}(\zeta) = \sum_{\ell=0}^{\infty} \mathcal{J}_{\ell} \zeta^{\ell} 
\end{equation} 
has a very simple form when rewritten in terms of the Miwa's shifts:
\begin{equation}
  \mathcal{J}(\zeta) 
  = 
  (\mySpd{\zeta}U) V. 
\label{cl-JUV} 
\end{equation}
The main result of this section is given by the following 
\begin{proposition}
\label{prop-cl}
The function $\mathcal{J}(\zeta)$ given by \eref{cl-JUV} satisfies 
equations 
\begin{subequations}
\begin{eqnarray}
  \left( \zeta - \xi\right) \partial(\xi)\mathcal{J}(\zeta)  
  & = & 
   \xi \;\partial_{1}\mathcal{K}(\zeta,\xi)  
\label{cl-pos} 
\\	
  \left( 1- \zeta \eta \right) \bar\partial(\eta)\mathcal{J}(\zeta)  
  & = & 
  \partial_{1}\mathcal{L}(\zeta,\eta)  
\label{cl-neg} 
\end{eqnarray}
\end{subequations}
where
\begin{eqnarray}
  \mathcal{K}(\zeta,\xi) 
  & = & 
  H(\xi) \left[ 1+ \zeta (\mySpd{\zeta}U) (\mySpi{\xi}V) \right]  
\label{cl:1902} 
\\	
  \mathcal{L}(\zeta,\eta) 
  & = & 
  h(\eta) \left[ 1- \zeta \eta (\mySpd{\zeta}U) (\mySni{\eta}v) \right]  
\end{eqnarray}
\end{proposition}
\noindent
whose proof is outlined in \ref{proof-cl}.

Expanding $\mathcal{K}(\zeta,\xi)$ and $\mathcal{L}(\zeta,\eta)$ in the double 
series, 
\begin{eqnarray}
  \frac{ \xi }{ \zeta - \xi } \, 
  \mathcal{K}(\zeta,\xi)  
  & = & 
  \sum_{\ell=0}^{\infty} \sum_{j=1}^{\infty} 
  \mathcal{K}_{\ell j} \zeta^{\ell}\xi^{j} 
\\	
  \frac{ 1 }{ 1 - \zeta\eta }\, 
  \mathcal{L}(\zeta,\eta)  
  & = & 
  1 + 
  \sum_{\ell=0}^{\infty} \sum_{k=1}^{\infty} 
  \mathcal{L}_{\ell k} \zeta^{\ell}\eta^{k} 
\end{eqnarray}
one arrives at local conservation laws for all `times', 
both positive and negative, 
\begin{eqnarray}
  \partial_{j} \mathcal{J}_{\ell} 
  & = & 
  \partial_{1} \mathcal{K}_{\ell j}, 
\\	
  \bar\partial_{k} \mathcal{J}_{\ell} 
  & = & 
  \partial_{1} \mathcal{L}_{\ell k} 
\end{eqnarray}
for $\ell=0,1,...$ and $j,k=1,2,...$ 
with $t_{1}$ being the distinguished variable.

\section{Bilinearization. \label{sec-bilin}}

The first step of the bilinearization of the problem is to introduce the 
tau-functions, which can be done in a standard way:

\begin{equation}
  U = \frac{ \sigma }{ \tau }, 
  \qquad
  V = \frac{ \tilde\rho }{ \tilde\tau } 
\end{equation}
and 
\begin{equation}
  u = \frac{ \tilde\sigma }{ \tilde\tau }, 
  \qquad
  v = \frac{ \rho }{ \tau }. 
\end{equation}
The second (and less trivial) step is to present the potential $\Psi$ as
$ \Psi = \tilde\tau / \tau $ which implies
\begin{equation}
  G(\xi) 
  = 
  \frac{\tau (\mySpd{\xi}\tilde\tau)}{(\mySpd{\xi}\tau) \tilde\tau},
  \qquad
  g(\eta) 
  = 
  \frac{ (\mySnd{\eta}\tau) \tilde\tau }{ \tau (\mySnd{\eta}\tilde\tau)}. 
\end{equation}
By a simple algebra one can show that equations \eref{pn-miwa} as well as 
definitions \eref{gp-def} and \eref{gn-def} become bilinear, which is
demonstrated by the following 
\begin{proposition}
\label{prop-hi-miwa}
The bilinear form of the extended DNLSH is given by the systems 
\begin{subequations}
\label{hi-miwa-p}
\begin{eqnarray}
  0
  & = & 
  \xi (\mySpd{\xi}\sigma) \tilde\tau 
  - (\mySpd{\xi}\tilde\sigma) \tau 
  + \tilde\sigma (\mySpd{\xi}\tau) 
\\
  0
  & = & 
  \xi \tilde\rho (\mySpd{\xi}\tau) 
  - \rho (\mySpd{\xi}\tilde\tau) 
  + (\mySpd{\xi}\rho) \tilde\tau 
\\
  0
  & = & 
  \xi \tilde\rho (\mySpd{\xi}\sigma) 
  - \tau (\mySpd{\xi}\tilde\tau) 
  + (\mySpd{\xi}\tau) \tilde\tau 
\label{hi-p-tau} 
\end{eqnarray}
\end{subequations}
(the positive subhierarchy) and 
\begin{subequations}
\label{hi-miwa-n}
\begin{eqnarray}
  0
  & = & 
  \eta (\mySnd{\eta}\tilde\sigma) \tau 
  - (\mySnd{\eta}\sigma) \tilde\tau 
  + \sigma (\mySnd{\eta}\tilde\tau) 
\\
  0
  & = & 
  \eta \rho (\mySnd{\eta}\tilde\tau) 
  - \tilde\rho (\mySnd{\eta}\tau) 
  + (\mySnd{\eta}\tilde\rho) \tau 
\\
  0
  & = & 
  \eta \rho (\mySnd{\eta}\tilde\sigma) 
  - (\mySnd{\eta}\tau) \tilde\tau 
  + \tau (\mySnd{\eta}\tilde\tau) 
\end{eqnarray}
\end{subequations}
(the negative one).
\end{proposition}

\noindent
From the practical viewpoint, this form of the DNLSH is most suitable when 
one wants to derive explicit solutions. 
In section \ref{sec-dark} we show 
how to obtain from \eref{hi-miwa-p} and \eref{hi-miwa-n} soliton solutions  
for the DNLSH by very simple calculations. 
However, for the sake of completeness, we show below the more traditional 
representation of the DNLSH in terms of the Hirota operators $D(\xi)$ and 
$\bar{D}(\eta)$ defined by  
\begin{equation}
  D(\xi) = \sum_{j=1}^{\infty} \xi^{j}\bar{D}_{j},
  \qquad 
  \bar{D}(\eta) = \sum_{k=1}^{\infty} \eta^{k}\bar{D}_{k} 
\end{equation}
where
\begin{eqnarray}
  D_{j} \, a \cdot b & = & 
  \left( \partial_{j} a \right) b - a \left( \partial_{j} b \right), 
\\
  \bar{D}_{k} \, a \cdot b & = & 
  \left( \bar\partial_{k} a \right) b - a \left( \bar\partial_{k} b \right) 
\end{eqnarray}
with 
$\partial_{j} = \partial / \partial t_{j}$ 
and 
$\bar\partial_{k} = \partial / \partial \bar{t}_{k}$. 
To do this, we need the two-shift generalizations of equations from 
Proposition \ref{prop-hi-miwa}, which we do not present here. Then, by taking 
the limits (as in \eref{def-part-pos}) one can obtain 
\begin{equation}
  H(\xi) 
  = 
  \myTheta_{\xi} 
  \frac{ (\mySpd{\xi}\tau) (\mySpi{\xi}\tilde\tau) }{ \tau \tilde\tau }, 
  \qquad
  h(\eta)  = 
  \mytheta_{\eta} 
  \frac{ (\mySni{\eta}\tau) (\mySnd{\eta}\tilde\tau) }{ \tau \tilde\tau } 
\end{equation}
(where $\myTheta_{\xi}$ and $\mytheta_{\eta}$ are constants that depend on 
$\xi$, $\eta$ and the boundary conditions) and the following representation 
of the DNLSH:
\begin{proposition}
\label{prop-hi-diff} 
The Hirota-like form of the extended DNLSH is given by the systems 
\begin{subequations}
\label{hi-part-p}
\begin{eqnarray}
  i \;D(\xi) \, \tau \cdot\tilde\tau   
  & = &
  - \tau \tilde\tau 
  + \myTheta_{\xi} (\mySpd{\xi}\tau) (\mySpi{\xi}\tilde\tau) 
\\
  i \;D(\xi) \, \sigma \cdot\tilde\tau   
  & = & 
  - \sigma \tilde\tau 
  + \myTheta_{\xi} (\mySpd{\xi}\sigma) (\mySpi{\xi}\tilde\tau) 
\\
  i \;D(\xi) \, \tilde\rho \cdot\tau   
  & = & 
  \phantom{-} 
  \tilde\rho \tau 
  - \myTheta_{\xi} (\mySpi{\xi}\tilde\rho) (\mySpd{\xi}\tau) 
\\
  i \;D(\xi) \, \tilde\sigma \cdot\tau   
  & = & 
  \phantom{-} 
  \myTheta_{\xi} \xi (\mySpd{\xi}\sigma) (\mySpi{\xi}\tilde\tau) 
\\
  i \;D(\xi) \, \rho \cdot\tilde\tau   
  & = & 
  - \myTheta_{\xi} \xi (\mySpi{\xi}\tilde\rho) (\mySpd{\xi}\tau) 
\end{eqnarray}
\end{subequations}
(the positive subhierarchy) and 
\begin{subequations}
\label{hi-part-n}
\begin{eqnarray}
  i \;\bar{D}(\eta) \, \tau \cdot\tilde\tau   
  & = & 
  \phantom{-} 
  \tau \tilde\tau 
  - \mytheta_{\eta} (\mySni{\eta}\tau) (\mySnd{\eta}\tilde\tau) 
\\
  i \;\bar{D}(\eta) \, \tilde\sigma \cdot\tau   
  & = & 
  - \tilde\sigma \tau 
  + \mytheta_{\eta} (\mySnd{\eta}\tilde\sigma) (\mySni{\eta}\tau) 
\\
  i \;\bar{D}(\eta) \, \rho \cdot\tilde\tau   
  & = & 
  \phantom{-} 
  \rho \tilde\tau 
  - \mytheta_{\eta} (\mySni{\eta}\rho) (\mySnd{\eta}\tilde\tau) 
\\
  i \;\bar{D}(\eta) \, \sigma \cdot\tilde\tau   
  & = & 
  \phantom{-} 
  \mytheta_{\eta} \eta (\mySnd{\eta}\tilde\sigma) (\mySni{\eta}\tau) 
\\
  i \;\bar{D}(\eta) \, \tilde\rho \cdot\tau   
  & = & 
  - \mytheta_{\eta} \eta (\mySni{\eta}\rho) (\mySnd{\eta}\tilde\tau) 
\end{eqnarray}
\end{subequations}
(the negative one).
\end{proposition}

\section{Solitons of the extended DNLSH under non-vanishing boundary conditions. 
\label{sec-dark}}

In this section we present the soliton solutions for the extended (describing 
both positive and negative flows) DNLSH under non-vanishing boundary conditions. 
Calculations that lead to our goal are similar to ones described in \cite{V12}. 
Thus, here we only outline the main steps and present the main results.

The building blocks for the soliton solutions are $N \times N$ matrices 
$\mymatrix{A}$ that satisfy the `almost rank-one' condition 
\begin{equation}
\label{ds-A}
  \mymatrix{L} \mymatrix{A} 
  - 
  \mymatrix{A} \mymatrix{L}^{-1} 
  = 
  | \,\ell\, \rangle \langle a | 
\end{equation}
where $\mymatrix{L}$ is a constant diagonal matrix, 
$| \,\ell\, \rangle$ is a constant $N$-component column, 
$| \,\ell\, \rangle = \left( \ell_{1}, ... , \ell_{N} \right)^{T}$, 
$\langle a |$ is a $N$-component row depending on the coordinates 
describing the DNLSH flows,
$\langle a\left( \mathrm{t}, \bar{\mathrm{t}} \right) | = 
\left( 
  a_{1}\left( \mathrm{t}, \bar{\mathrm{t}} \right), ... , 
  a_{N}\left( \mathrm{t}, \bar{\mathrm{t}} \right) 
\right)$, 
and matrices $\mymatrix{H}_{\zeta}$ that are defined by 
\begin{equation}
  \mymatrix{H}_{\zeta} 
  = 
  \left( \zeta \mymatrix{1} - \mymatrix{L} \right) 
  \left( \zeta \mymatrix{1} - \mymatrix{L}^{-1} \right)^{-1}
\end{equation}
where $\mymatrix{1}$ is the $N \times N$ unit matrix.
The remarkable property of the above matrices, that is used 
below, is that the determinants
\begin{equation}
  \omega\left( \mymatrix{A} \right) 
  = 
  \det \left| \mymatrix{1} + \mymatrix{A} \right|
\label{omega-def}
\end{equation}
satisfy the Fay-like identity
\begin{equation} 
  (\xi   - \eta)  \, \omega_{\zeta} \, \omega_{\xi\eta} 
  + 
  (\eta  - \zeta) \, \omega_{\xi}   \, \omega_{\eta\zeta} 
  +  
  (\zeta - \xi)   \, \omega_{\eta}  \, \omega_{\zeta\xi} 
  = 0 
\label{fay}
\end{equation}
where 
\begin{equation}
  \omega = \omega\left( \mymatrix{A} \right),
  \qquad
  \omega_{\zeta} = \omega\left( \mymatrix{A} \mymatrix{H}_{\zeta} \right),
  \qquad
  \omega_{\xi\eta} = 
  \omega\left( \mymatrix{A} \mymatrix{H}_{\xi} \mymatrix{H}_{\eta} \right).
\end{equation}

\begin{figure}%
\begin{center}
\begin{picture}(125,125)(0,0)
\put(25, 25)  {\circle{15}} \put(22, 22) {$\tilde\rho$}
\put(75, 25)  {\circle{15}} \put(72, 22) {$\tilde\tau$}
\put(125,25)  {\circle{15}} \put(122,22) {$\tilde\sigma$}
\put(25, 75)  {\circle{15}} \put(22, 72) {$\rho$}
\put(75, 75)  {\circle{15}} \put(72, 72) {$\tau$}
\put(125,75)  {\circle{15}} \put(122,72) {$\sigma$}
\put(32,25){\vector(1,0){35}} \put(45,15){$\mymatrix{H}_{\mu}$}
\put(82,25){\vector(1,0){35}} \put(95,15){$\mymatrix{H}_{\mu}$}
\put(32,75){\vector(1,0){35}} \put(45,65){$\mymatrix{H}_{\mu}$}
\put(82,75){\vector(1,0){35}} \put(95,65){$\mymatrix{H}_{\mu}$}
\put(23,32){\vector(0,1){35}} \put(27,45){$\mymatrix{H}_{1/\mu}$}
\put(73,32){\vector(0,1){35}} \put(77,45){$\mymatrix{H}_{1/\mu}$}
\put(123,32){\vector(0,1){35}} \put(127,45){$\mymatrix{H}_{1/\mu}$}
\end{picture}
\end{center}
\caption{Relations between the tau-functions.}
\label{fig-one}
\end{figure}
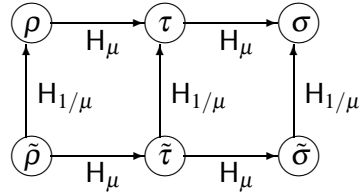

To obtain the soliton solutions, first one needs to establish the 
relationships between six tau-functions. 
This can be done by means of matrices  
$\mymatrix{H}_{\mu}$ and $\mymatrix{H}_{1/\mu}$, 
related by 
\begin{equation}
  \mymatrix{H}_{\mu} 
  \mymatrix{H}_{1/\mu} 
  = 
  \mymatrix{L}^{2} 
\end{equation}
as follows (see Fig. \ref{fig-one}): 
\begin{equation}
\label{ds-tau-0} 
  \tau = \tau_{*}\, 
    \omega\left( \mymatrix{A} \mymatrix{H}_{1/\mu} \right), 
  \qquad
  \sigma = \sigma_{*}\, 
    \omega\left( \mymatrix{A} \mymatrix{H}_{\mu}\mymatrix{H}_{1/\mu} \right), 
  \qquad 
  \rho = \rho_{*}\, 
    \omega\left( \mymatrix{A} \mymatrix{H}_{\mu}^{-1}\mymatrix{H}_{1/\mu} \right) 
\end{equation}
and
\begin{equation}
\label{ds-tau-1} 
  \tilde\tau = \tilde\tau_{*}\, 
    \omega\left( \mymatrix{A} \right), 
  \qquad
  \tilde\sigma = \tilde\sigma_{*}\, 
    \omega\left( \mymatrix{A} \mymatrix{H}_{\mu} \right), 
  \qquad 
  \tilde\rho = \tilde\rho_{*}\, 
    \omega\left( \mymatrix{A} \mymatrix{H}_{\mu}^{-1} \right) 
\end{equation}
The functions with the $*$-subscript (the background tau-functions)
with simple transformation properties with respect to the Miwa shifts 
(they are eigenfunctions of $\mySpd{\xi}$ and $\mySnd{\eta}$) 
are presented below. 

Secondly, one has to define the `evolution' of the matrices $\mymatrix{A}$ with 
respect to the Miwa shifts which, again, can be constructed by means of the 
$\mymatrix{H}$-matrices,
\begin{equation}
  \mySpd{\xi}\mymatrix{A} 
  = 
  \mymatrix{A} \mymatrix{H}_{\alpha(\xi)} \mymatrix{H}_{\alpha(0)}^{-1}, 
  \qquad  
  \mySnd{\eta}\mymatrix{A} 
  = 
  \mymatrix{A} \mymatrix{H}_{\beta(\eta)} \mymatrix{H}_{\beta(0)}^{-1} 
\label{ds-evol}
\end{equation}
where $\alpha(\xi)$ and $\beta(\eta)$ should be determined (see below).  

Finally, one has to find the background tau-functions 
$\tau_{*}$, $\sigma_{*}$, $\rho_{*}$, 
$\tilde\tau_{*}$, $\tilde\sigma_{*}$ and $\tilde\rho_{*}$. 
This can be done by substituting \eref{ds-tau-0} and \eref{ds-tau-1} into 
the Fay identities and gathering all the constants. 
For example, equation \eref{fay} with $(\xi,\eta,\zeta)=(0,1/\mu,\alpha)$ after 
the shift $\mymatrix{A} \to \mymatrix{A}\mymatrix{H}_{0}^{-1}$
and replacement of all 
$\omega\left( \mymatrix{A} \mymatrix{H}_{\zeta} \right)$ with the corresponding 
tau-functions from \eref{ds-tau-0}, \eref{ds-tau-1} and \eref{ds-evol} becomes 
(under the assumption $\alpha(0)=0$)
\begin{equation}
  \mu\alpha 
  \frac{ \tilde\rho }{ \tilde\rho_{*} }
  \frac{ (\mySpd{\xi} \sigma) }{ (\mySpd{\xi} \sigma_{*}) }
  - 
  \frac{ \tau }{ \tau_{*} }
  \frac{ (\mySpd{\xi} \tilde\tau) }{ (\mySpd{\xi} \tilde\tau_{*}) }
  + 
  (1 - \mu\alpha)
  \frac{ \tilde\tau }{ \tilde\tau_{*} }
  \frac{ (\mySpd{\xi} \tau) }{ (\mySpd{\xi} \tau_{*}) }
  = 0.
\end{equation}
Comparing this identity with \eref{hi-p-tau} one can conclude that to solve the latter 
one has to met the conditions
\begin{equation}
  \frac{1}{\xi}
  \frac{ \mu\alpha(\xi) }{ \tilde\rho_{*} (\mySpd{\xi} \sigma_{*}) }
  = 
  \frac{ 1 }{ \tau_{*} (\mySpd{\xi} \tilde\tau_{*}) }
  = 
  \frac{ 1 - \mu\alpha(\xi) }{ \tilde\tau_{*} (\mySpd{\xi} \tau_{*}) }.
\label{ds-eq-star}
\end{equation}
Repeating this proceeding with different choices of $(\xi,\eta,\zeta)$ one can 
reduce \eref{hi-miwa-p} to a system of equations similar to \eref{ds-eq-star} 
that can be solved by the proper choice of the background tau-functions and 
$\alpha(\xi)$. This leads to the following results.

\begin{proposition}
\label{prop-ds-hi-p}
Functions \eref{ds-tau-0}, \eref{ds-tau-1} 
solve equations (\ref{hi-miwa-p}) provided 
the background tau-functions satisfy
\begin{equation}
  \begin{array}{lcl}
  \tau_{*}^{2} 
  & = & 
  \left( 1 - \mu^{-2} \right) 
  \sigma_{*}\rho_{*}, 
  \\[1mm]
  \tilde\tau_{*}^{2} 
  & = & 
  \left( 1 - \mu^{-2} \right) 
  \tilde\sigma_{*}\tilde\rho_{*} 
  \end{array}
\label{ds-restriction}  
\end{equation}
while their dependence on positive times 
is governed by 
\begin{equation}
  \begin{array}{lcl}
  (\mySpd{\xi}\tau_{*}) \tilde\tau_{*} 
  & = & 
  \tau_{*} (\mySpd{\xi}\tilde\tau_{*}) \chi_{\tau}(\xi), 
  \\[1mm]
  (\mySpd{\xi}\sigma_{*}) \tau_{*} 
  & = & 
  \sigma_{*} (\mySpd{\xi}\tau_{*}) \chi_{\sigma}(\xi) 
  \\[1mm]
  (\mySpd{\xi}\tilde\sigma_{*}) \tilde\tau_{*} 
  & = & 
  \tilde\sigma_{*} (\mySpd{\xi}\tilde\tau_{*}) \chi_{\sigma}(\xi), 
  \end{array}
\label{ds-miwa-p}  
\end{equation}
where functions $\chi_{\tau}(\xi)$ and $\chi_{\sigma}(\xi)$ are given by 
\begin{equation}
\begin{array}{lcl}
  \chi_{\tau}(\xi) & = & 1 - \mu\alpha(\xi), 
  \\[1mm]
  \chi_{\sigma}(\xi) & = & 1 - \mu^{-1}\alpha(\xi) 
\end{array}
\end{equation}
and $\alpha(\xi)$ is the solution of 
\begin{equation}
  \alpha(\xi) + \alpha(\xi)^{-1} 
  = 
  \mu + \mu^{-1} 
  + c\mu \xi^{-1},
  \qquad
  \alpha(0) = 0
\label{ds-law-p}
\end{equation}
where constant $c$ determines the amplitude of the solutions, 
\begin{equation}
  c 
  = 
  \frac{ \tau_{*}\tilde\tau_{*} }{ \sigma_{*}\tilde\rho_{*} } 
  =
  \left( 1 - \mu^{-2} \right)^{2}
  \frac{ \tilde\sigma_{*}\rho_{*} }{ \tau_{*}\tilde\tau_{*} }. 
\label{ds-ampl}
\end{equation}
\end{proposition} 
\noindent
(The dependence of $\rho_{*}$ and $\tilde\rho_{*}$ on positive times, which is not 
written explicitly, can be recovered from \eref{ds-miwa-p} and 
\eref{ds-restriction}.)

Similar result can be obtained for the negative subhierarchy. 

\begin{proposition}
\label{prop-ds-hi-n}
Functions \eref{ds-tau-0}, \eref{ds-tau-1} 
solve equations \eref{hi-miwa-p} provided 
the background tau-functions satisfy \eref{ds-restriction}  
while their dependence on negative times 
is governed by 
\begin{equation}
  \begin{array}{lcl}
  (\mySnd{\eta}\tau_{*}) \tilde\tau_{*} 
  & = & 
  \tau_{*} (\mySnd{\eta}\tilde\tau_{*}) {\bar\chi}_{\tau}(\eta), 
  \\[1mm]
  (\mySnd{\eta}\sigma_{*}) \tau_{*} 
  & = & 
  \sigma_{*} (\mySnd{\eta}\tau_{*}) {\bar\chi}_{\sigma}(\eta), 
  \\[1mm]
  (\mySnd{\eta}\tilde\sigma_{*}) \tilde\tau_{*} 
  & = & 
  \tilde\sigma_{*} (\mySnd{\eta}\tilde\tau_{*}) {\bar\chi}_{\sigma}(\eta) 
  \end{array}
\end{equation}
where functions ${\bar\chi}_{\tau}(\eta)$ and ${\bar\chi}_{\sigma}(\eta)$ are given by 
\begin{equation}
  \begin{array}{lcl}
  {\bar\chi}_{\tau}(\eta) 
  & = & \displaystyle
  \frac{ 1 - \mu\beta(\eta) }{ 1 - \mu^{2} },
  \\[3mm]
  {\bar\chi}_{\sigma}(\eta) 
  & = & \displaystyle
  \frac{ \beta(\eta) }{ \mu } \, 
  \frac{ 1 - \mu^{2} }{ 1 - \mu\beta(\eta) }
  \end{array}
\end{equation}
and $\beta(\eta)$ is the solution of 
\begin{equation}
  \beta(\eta) + \beta(\eta)^{-1} 
  = 
  \mu + \mu^{-1} 
  + c\mu \eta,
  \qquad
  \beta(0) = \mu.
\label{ds-law-n}
\end{equation}
\end{proposition}

Knowing solutions for the bilinear system \eref{hi-miwa-p} and 
\eref{hi-miwa-n} it is easy to write down solutions for \eref{pn-miwa}. 
Bearing in mind physical applications, one has to restrict himself with 
$U = \varepsilon V^{*}$ where the asterisk stands for the complex conjugation 
and $\varepsilon = \pm 1$. It should be noted that, contrary to the case of 
the nonlinear Schr\"odinger equation where the value of $\varepsilon$ is 
crucial (it determines, for example, the type of solitons, bright or dark), 
$\varepsilon$ can be eliminated from the DNLS equation by the substitution 
$U(t_{1},t_{2},...) \to U(\varepsilon t_{1},t_{2},...)$. Thus, we consider 
below only the case 
\begin{equation}
\label{ds-involution}
  U = V^{*}.
\end{equation}
This condition imposes some restrictions on the soliton matrices 
$\mymatrix{A}$ and the transformation matrices $\mymatrix{H}$. 
It can be shown that to resolve \eref{ds-involution} one has to ensure the 
reality of $\mymatrix{C}=\mymatrix{A}\mymatrix{H}_{1/\mu}^{1/2}$, 
$\mymatrix{C}^{*}=\mymatrix{C}$, and unitarity 
of all $\mymatrix{H}$-matrices used above:
\begin{equation}
  \mymatrix{H}_{\zeta}\mymatrix{H}_{\zeta}^{*} = \mymatrix{1}, 
  \qquad
  \zeta = \mu, \, 1/\mu, \, \alpha(\xi), \, \beta(\eta)
\end{equation}
for real $\xi$ and $\eta$.
It is easy to verify that one can met these conditions by taking real $\mu$ 
and 
\begin{equation}
  \mymatrix{L} = \mbox{diag} \left( e^{i\theta_{n}} \right)_{n = 1, ..., N} 
\end{equation}
which after omitted here calculations leads to 
\begin{equation}
  \mymatrix{C}\left( \mathrm{t}, \bar{\mathrm{t}} \right) = 
  \left( 
    \mymatrix{C}^{(0)}_{mn} \, 
    e^{ \nu_{n}\left( \mathrm{t}, \bar{\mathrm{t}} \right) } 
  \right)_{m,n = 1, ..., N}  
\end{equation}
where $\left( \mymatrix{C}^{(0)}_{mn} \right)$ is a constant matrix and 
\begin{equation}
  \nu_{n}\left( \mathrm{t}, \bar{\mathrm{t}} \right) = 
  \sum\limits_{k=1}^{\infty} 
  \left( \nu_{nk} t_{k} + \tilde\nu_{nk} \bar{t}_{k} \right) 
\label{ds-nu-def}
\end{equation}
with 
\begin{subequations}
\label{ds-nu-series}
\begin{eqnarray}
  \sum\limits_{k=1}^{\infty} \nu_{nk} \xi^{k} / k  
  & = & 
  2 \arg \left[ 1 - \alpha(\xi) e^{-i\theta_{n}} \right], 
  \\ 
  \sum\limits_{k=1}^{\infty} \tilde\nu_{nk} \eta^{k} / k  
  & = & 
  2 \arg \left[ 1 - \beta(\eta) e^{-i\theta_{n}} \right]   
  - 2 \arg \left[ 1 - \mu e^{-i\theta_{n}} \right]. 
\end{eqnarray}
\end{subequations}
The ratio of the background tau-functions, $U_{*}=\sigma_{*}/\tau_{*}$ can be 
presented as 
\begin{equation}
  U_{*} = U^{(0)} 
  e^{ i\varphi\left( \mathrm{t}, \bar{\mathrm{t}} \right) } 
\end{equation}
with arbitrary constant $U^{(0)}$ that replaces $c$ in 
\eref{ds-law-p}, \eref{ds-ampl}, \eref{ds-law-n}, 
\begin{equation}
  c = \bigl| U^{(0)} \bigr|^{-2} 
\end{equation}
and 
\begin{equation}
  \varphi\left( \mathrm{t}, \bar{\mathrm{t}} \right) =  
  \sum_{k=1}^{\infty} 
  \left( \varphi_{k} t_{k} + \tilde\varphi_{k} \bar{t}_{k}\right) 
\label{ds-phi-def}
\end{equation}
where
\begin{subequations}
\label{ds-phi-series}
\begin{eqnarray}
  \sum\limits_{k=1}^{\infty} \varphi_{k} \xi^{k} / k  
  & = & 
  - \ln \left[ 1 - \mu^{-1} \alpha(\xi) \right], 
\\ 
  \sum\limits_{k=1}^{\infty} \tilde\varphi_{k} \eta^{k} / k 
  & = & 
  \ln \left[ 1 - \mu^{-1} \beta(\eta)^{-1} \right]
  - \ln \left[ 1 - \mu^{-2} \right]. 
\end{eqnarray}
\end{subequations}
Finally, presenting 
$\mymatrix{H}_{\mu}\mymatrix{H}_{1/\mu}^{1/2}$
and 
$\mymatrix{H}_{1/\mu}^{1/2}$ 
as
$\mbox{diag}\left( e^{i \gamma_{n}^{(1,2)}} \right)$ 
where
\begin{equation}
  \gamma_{n}^{(2)} = \arg \left( 1 - \mu e^{i\theta_{n}} \right), 
  \qquad 
  \gamma_{n}^{(1)} = 2\theta_{n} - \gamma_{n}^{(2)}, 
\label{ds-gamma-def}
\end{equation}
using \eref{ds-A} to obtain $\mymatrix{C}^{(0)}_{mn}$ 
and eliminating, without loss of generality,  superfluous constants upon 
noting that the determinants \eref{omega-def} are invariant under 
transformations $\mymatrix{A} \to \mymatrix{M}^{-1} \mymatrix{A} \mymatrix{M}$, 
one arrives at the final expressions for the soliton solutions of the 
extended DNLSH:
\begin{proposition}
\label{prop-ds-sol}
The $N$-soliton solutions for the DNLSH under non-vanishing boundary conditions 
can be presented as
\begin{equation}
  U\left( \mathrm{t}, \bar{\mathrm{t}} \right) 
  = 
  U^{(0)}
  \exp\left[ i\varphi\left( \mathrm{t}, \bar{\mathrm{t}} \right) \right] 
  \frac{ \Delta_{1}\left( \mathrm{t}, \bar{\mathrm{t}} \right) }
       { \Delta_{2}\left( \mathrm{t}, \bar{\mathrm{t}} \right) }
\end{equation}
with arbitrary $U^{(0)}$ and 
\begin{equation}
	\Delta_{\ell}\left( \mathrm{t}, \bar{\mathrm{t}} \right)
  = 
  \det\left| 
    \delta_{mn} + 
    C^{(0)}_{n} 
    \frac{ \exp\left[ \nu_{n}\left( \mathrm{t}, \bar{\mathrm{t}} \right) 
                      + i\gamma_{n}^{(\ell)} \right] }
         { \sin\left( \frac{ \theta_{m} + \theta_{n} }{ 2 } \right) } 
    \right|_{m,n =1, ..., N} 
  \qquad
  (\ell = 1,2).
\end{equation}
where functions 
$\nu\left( \mathrm{t}, \bar{\mathrm{t}} \right)$ 
and 
$\varphi\left( \mathrm{t}, \bar{\mathrm{t}} \right)$ 
are defined in \eref{ds-nu-def}, \eref{ds-nu-series} and 
\eref{ds-phi-def}--\eref{ds-gamma-def}, 
$C^{(0)}_{n}$ are arbitrary real constants and $V = U^{*}$.
\end{proposition}

\section{Conclusion.}

To conclude, we would like to summarize the main results of this paper and to 
outline possible continuations of the presented work.

The main subject of this paper is the extended DNLSH. The main results we have 
obtained are 1) the functional representation of both positive (classical) and 
negative flows, 2) the generating function for the conservation laws and 3) the 
dark-soliton solutions. 

The most straightforward continuation of this work is to use the advantages of 
the functional representation and to derive other classes of explicit solutions 
which complement the bright-soliton solutions derived in \cite{FGZ13} and 
dark solitons presented above. This can be done starting from the bilinear 
equations of proposition \ref{prop-hi-miwa} which can be associated with the 
Fay identities for the theta-functions and used to derive the quasiperiodic 
solutions, or with various determinant identities that lead to Wronskian, 
Toeplitz and other solutions. 

Another range of arising problems is related to the results of section 
\ref{sec-mix}. We would like to stress that this paper is not aimed to elaborate  
methods of generating new integrable models, but the examples presented in 
section \ref{sec-exm}, 
which surely do not exhaust all models `hidden' inside the DNLSH, demonstrate that 
the question of what systems can be obtained from (or reduced to) the equations of 
a given hierarchy (the DNLSH in our case) is far from trivial. 
Even in the short list of section \ref{sec-exm} one can find a few `new', 
i.e. not well-studied, ones: 
(1+2)-dimensional Chen-Lee-Liu equation \eref{ex-cll-21}, 
(1+2)-dimensional Kaup-Newell equation \eref{ex-kn-21}, 
Adler-Shabat $H_{5}$ system \eref{ex-as}.

Considering the last one, it was obtained by Adler and Shabat in the framework 
of the classification of two-component hyperbolic systems. However, the 
result of \cite{AS06} seems to be the only fact that we know about this model, 
while the typical set of questions related to any integrable model (the 
inverse scattering transform, conservation laws, explicit solutions) is still 
to be studied.

The (1+2)-dimensional systems \eref{ex-cll-21} and \eref{ex-kn-21} have been 
mentioned in the literature, see \cite{K91,S92,SY97,MY97,Z06,T11}. 
However, these papers are devoted mostly to the interrelations between various 
integrable models and the algebraic structures behind them. At the same time, 
the conservation laws or explicit solutions are, again, have not been derived 
yet. From this viewpoint the results presented above not only give some 
additional insight into the place of these equations among other integrable 
models but also give possibility of presenting a wide range of explicit 
solutions by modifying ones obtained for the DNLSH. 
For example, a corollary of proposition \ref{prop-ds-sol} is that it can be 
used to describe the dark solitons of \eref{ex-cll-21} and \eref{ex-kn-21}. 
To complete this task one has to `extract' from \eref{ds-nu-series} and 
\eref{ds-phi-series} the explicit dependence on the 
lowest `times' (indicated in \eref{ex-cll-times} and \eref{ex-kn-times}) and 
to write down proper combinations of the tau-functions. 

However, these questions are out of scope of this article and, to our opinion, deserve 
separate studies.


\appendix

\section{Proof of Proposition \label{proof-cl}}

The proof is straightforward (but rather tedious) application of 
\eref{pp-part-xi}, \eref{pn-part} together with 
\eref{pp-miwa}, \eref{pn-miwa}.
Below we outline the main steps leaving the details omitted.

First, starting from \eref{pp-part-xi} one can derive 
\begin{equation}
  i \partial(\xi)\mathcal{J}(\zeta)  
  = 
  \zeta^{-1} \left[ H(\xi) - \mySpd{\zeta}H(\xi) \right]  G(\zeta) 
\label{cl:1901} 
\end{equation}
that can be rewritten, with the help of \eref{pp-miwa}, as
\begin{equation}
  i \left( \xi - \zeta \right)  \;\partial(\xi)\mathcal{J}(\zeta)  
  = 
  H(\xi) A(\xi,\zeta) B(\xi,\zeta) 
\label{eq:1924} 
\end{equation}
where 
\begin{eqnarray}
  A(\xi,\zeta) 
  & = & 
  \xi G(\zeta) 
  - \zeta G(\xi), 
\\
  B(\xi,\zeta) 
  & = &
  (\mySpd{\xi\zeta}U) V 
  - (\mySpd{\zeta}U) (\mySpi{\xi}V). 
\end{eqnarray}
On the other hand, application of \eref{pp-part1} to the derivative 
of \eref{cl:1902} gives
\begin{equation}
  i \xi \;\partial_{1}\mathcal{K}(\zeta,\xi)  
  = 
  - A(\xi,\zeta) B(\xi,\zeta) H(\xi) 
\end{equation}
which proves \eref{cl-pos}.

Considering \eref{cl-neg}, its verification is based on the identity 
\begin{equation}
  \left[ 1- \zeta \eta (\mySpd{\zeta}U) (\mySni{\eta}v) \right]  h(\eta) 
  = 
  \left[ 1- \zeta \eta (\mySpd{\zeta}\mySnd{\eta}u) V \right]  (\mySpd{\zeta}h(\eta)) 
\label{cl-neg-key} 
\end{equation}
which is, again, a consequence of \eref{pn-miwa}. 
Differentiating \eref{cl-JUV} and using \eref{cl-neg-key} one can obtain 
an equation, which is similar to \eref{cl:1901}:
\begin{equation}
  i \;\bar\partial(\eta)\mathcal{J}(\zeta)  
  = 
  \zeta^{-1} \left[ \mySpd{\zeta}h(\eta) - h(\eta) \right]  G(\zeta). 
\label{cl-neg-11} 
\end{equation}
On the other hand, the expression for derivative $\partial_{1}\mathcal{L}$, 
that stems from the $\xi \to 0$ limit of \eref{eq:1811}, \eref{eq:1812} 
combined with \eref{pp-part1}, 
\begin{equation}
  i \partial_{1}\mathcal{L}(\zeta,\eta)  
  = 
  \eta \left[ 1- (\mySpd{\zeta}U) (\mySni{\eta}v) \right]  G(\zeta) h(\eta) 
  - \eta \left[ 1- (\mySnd{\eta}u) V \right]  \mathcal{L}(\zeta,\eta) 
\end{equation}
can be transformed, using again \eref{cl-neg-key}, into   
\begin{equation}
  i \;\partial_{1}\mathcal{L}(\zeta,\eta)  
  = 
  \eta \left[ 1- (\mySpd{\zeta}U) (\mySni{\eta}v) \right]  G(\zeta) h(\eta) 
  - \eta \left[ 1- (\mySpd{\zeta}\mySnd{\eta}u) V \right]  G(\zeta) (\mySpd{\zeta}h(\eta)) 
\label{cl-neg-25} 
\end{equation}
from which equation \eref{cl-neg} follows immediately.

\end{document}